\documentclass[11pt]{article}

\usepackage{epsf,epsfig,graphicx}

\def\be{\begin{eqnarray}}
\def\en{\end{eqnarray}}
\def\non{\nonumber}
\def\ra{\rangle}

\def\sl{\!\!\!\slash}
\def\pr{{Phys. Rev.}~}

\def\pl{{ Phys. Lett.}~}
\def\np{{ Nucl. Phys.}~}
\def\epj{{ Eur. Phys. J.}~}

\begin{document}

\title{Study of scalar mesons $f_0(980)$ and $f_0(1500)$ from
$B \to f_0(980) K$ and $B \to f_0(1500) K$ Decays }

\author{Wei Wang$^b$,
Yue-Long Shen$^b$,
Ying Li$^b$ and Cai-Dian L\"u$^{a,b}$\\
{\it \small $a$   CCAST (World Laboratory), P.O. Box 8730,
Beijing 100080, China;}\\
{\it \small $b$    Institute of High Energy Physics, CAS, P.O.Box
918(4) } {\it \small Beijing 100049, China}\footnote{Mailing
address.}}

\maketitle

\begin{abstract}
Within perturbative QCD approach based on $k_T$ factorization, we
analyze the scalar mesons $f_0(980)$ and $f_0(1500)$ productions
in $B$ decays. By identifying $f_0(980)$ as the composition of
$\bar ss$ and $\bar nn=(\bar uu+\bar dd)/\sqrt2$, we calculate the
exclusive decays $B\to f_0(980) K$. We find that the
non-factorization $f_0$-emission diagrams can give larger
contribution to the branching ratio, than the previous PQCD
calculation. Our new results can explain the current experimental
data well. Under the assumption of quarkonium dominance, we study
the branching ratio of decays $B\to f_0(1500) K$. The results show
that in the two-quark picture of $f_0$ meson the contribution from
$\bar ss$ component is at the similar size as that from the $\bar
n n$ component. Comparing with the data, our results show the
preference of $f_0(1500)$ as a member of the ground state of
scalar $\bar qq$ nonet. Similar results can also apply to
$f_0(1370)$ and $f_0(1710)$, if these mesons are dominated by the
quarkonium content. With more experimental data in future, these
studies will help us understand the intrinsic characters of these
scalar mesons.
\end{abstract}

\section{Introduction}\label{intro}

In spite of the striking success of QCD theory for strong
interaction, the underlying structure of the light scalar mesons
is still under controversy theoretically \cite{Spanier,Close}. In
the literature, there are many proposals such as $\bar qq$, $\bar
q\bar qqq$,   meson-meson bound states or even supplemented with a
scalar glueball. It is very likely that they are not made of one
simple component but are the superpositions of these contents and
it is model dependent to determine the dominant component. The
different scenarios may give very different predictions on the
production and decay of the scalar mesons which can be tested by
the related experiments. Although intensive study has been given
to the decay property of the scalar mesons, the production of
these mesons can provide a different unique insight to the
mysterious structure of these mesons, especially their production
in $B$ decays. Compared with $D$ meson decays, the role of scalar
particles in $B$ decays is much more noticeable because of the
larger phase space.

$f_0(980)$ is the first scalar meson observed in $B$ decays with
the decay mode $B\to f_0(980)K$. In the three-body decays
$B^\pm\to K^\pm\pi^\mp\pi^\pm$ \cite{Bellef0}, Belle found a large
branching ratio for $B^- \to K^-f_0(980)\to K^-(\pi^+\pi^-)$,
which was confirmed by BaBar \cite{BaBarf0} later. Using the
branching fraction of $f_0\to \pi^+\pi^-$, we can obtain a large
branching ratio at order $10^{-5}$ for the decay $B\to f_0(980)K$.
These measurements of the decay $B\to f_0(980)K$ has arisen much
interest on theoretical side.  The earlier Perturbative QCD (PQCD)
approach calculation \cite{Chenf0K1,Chenf0K2} shows a smaller
branching ratio than the experimental data. Recently, these modes
have also been studied to probe the new physics beyond standard
model in \cite{GMM} using the generalized factorization approach.
They find that in standard model, the branching ratio is quite
below the experimental values and therefore, these modes can be a
probe of the R-parity violation supersymmetric model. Within the
framework of QCD factorization approach (QCDF), $B\to f_0(980)K$
has  also been studied recently \cite{CYf0K,CCYscalar}. With the
decay constants and light-cone distribution amplitudes derived
from the QCD sum rules, the updated results in QCDF
\cite{CCYscalar} suffice to explain the experimental data. It is
necessary to re-analyze of these decay channels in PQCD in order
to find out whether the differences arise from the difference
between the two approaches or only the different non-perturbative
inputs.

For $B\to f_0(1500)K$, there is a puzzle in experiments: both
Belle \cite{BelleKpipi} and BaBar \cite{BaBarKKK} found a
resonance in the $K^+K^-$ mass spectrum of $B\to (K^+K^-)K$
decays, whose mass and width is consistent with $f_0(1500)$.  Due
to the large ratio  $\Gamma(f_0(1500)\to
\pi\pi)/\Gamma(f_0(1500)\to K\bar K)=4.06$ \cite{PDG}, we expect
the similar peak in the corresponding $\pi\pi$ channel. But there
is no signal in  the decays of $B\to K(\pi^+\pi^-)$
\cite{BelleKpipi,BaBarKpipi}. In order to make it clear, more
experimental data are required. On the other side we should also
know the theoretical predictions  on $B\to f_0(1500)K$.

Minkowski and Ochs \cite{MOBtoscalar} have studied $B\to
f_0(1500)K$ by the assumption that $f_0(1500)$ is pictured as the
superposition of $\bar ss$ and $\bar nn$ \cite{MOglue}. In their
study, the decay amplitudes are dominated by the QCD penguin
operators $b\to s\bar qq$ (q=u,d,s) and the chromomagnetic penguin
operator $O_{8g}$, while the tree operators are neglected for the
suppression of CKM matrix elements and annihilation topology
contribution is also omitted due to the power suppression of
$1/m_B$. But it is shown that the annihilation diagrams are not
negligible in $B\to\pi K$ etc. \cite{PQCD} which give a large
contribution to the imaginary part\footnote{In the recent study on
the factorization property for the annihilation contribution using
the effective theory \cite{ALRW}, the leading contributions of
order $\alpha_s(m_b)\Lambda_{QCD}/m_B$ are factorizable and
real.}. This implies the annihilation contribution may not be
negligible in $B\to f_0(1500)K$ either. In this paper we perform
the PQCD study on this decay mode to provide a systematic and
reliable analysis.

The outline of this paper is as follows: In Sect.\ref{proper}, we
briefly discuss the status of the study on the physical properties
of $f_0(980)$ and $f_0(1500)$. In Sect.\ref{results}, we calculate
the decays in PQCD approach with discussions. The final part
contains our conclusions.


\section{Physical properties of $f_0(980)$ and
$f_0(1500)$}\label{proper}

\subsection{Quark Structure}

 Although the quark model and QCD have achieved great
successes, the inner structure of scalar mesons is not well
established theoretically. There are many scenarios for the
classification of the scalar mesons. In the scheme proposed in
\cite{MOglue},  the $\sigma$ and $\kappa(900)$ are not considered
as physical states. The lowest scalar $\bar qq$ nonet is rather
formed by the iso-vector $a_0(980)$, the iso-scalars $f_0(980)$,
$f_0(1500)$ and the iso-doublet $K^*_0(1430)$. In the second
scheme, it has been suggested that the light scalars below or near
1 GeV-- $f_0(600)$ (or $\sigma$), $f_0(980)$, $K_0^*(800)$ (or
$\kappa$) and $a_0(980)$--form an $SU(3)$ flavor nonet, either
$\bar qq$ or $\bar q\bar q qq$, while scalar mesons above 1
{\mbox{ GeV}}, namely, $f_0(1370)$, $a_0(1450)$, $K^*_0(1430)$ and
$f_0(1500)/f_0(1710)$, form another nonet. According to the
different descriptions for the first nonet, this scheme is divided
into two different scenarios further, which we will denote as
scenario I and scenario II respectively in this work.

In scenario I, the first nonet is viewed as $\bar qq$ states. In
this scenario, $f_0(980)$ is mainly an $s\bar s$ state and this is
supported by the data of $D_s^+\to f_0\pi^+$ and $\phi\to
f_0\gamma$. However, there also exists some experimental evidences
indicating that $f_0(980)$ is not purely an $s\bar s$ state. The
observation of $\Gamma(J/\psi\to f_0\omega)\approx {1\over
2}\Gamma(J/\psi\to f_0\phi)$ \cite{PDG}   indicates the existence of
the non-strange and strange quark contents in $f_0(980)$. Therefore,
$f_0$ should be a mixture of $\bar ss$ and $\bar nn\equiv (\bar
uu+\bar dd)/\sqrt{2}$:
 \be
 |f_0(980)\ra = |s\bar s\ra\cos\theta+|n\bar n\ra\sin\theta,
 \en
with $\theta$ is the mixing angle. Experimental implications for the
mixing angle have been discussed in detail in Ref.
\cite{Chengmixing}: $\theta$ lies in the ranges of
$25^\circ<\theta<40^\circ$ and $140^\circ<\theta< 165^\circ$. In
scenario II, $f_0(980)$ is described by a four-quark state, which is
too complicated to be studied in a factorization approach. In order
to give quantitative predictions, we work in the scenario I for
$f_0(980)$ only and identifying it as the mixture of $\bar ss$ and
$\bar nn$.

In both scenario I and scenario II, $f_0(1500)$ could be treated
as a $\bar qq$ state, either the first-excited state or the ground
state. But the case becomes complicated by the possible existence
of glueball content. Glueball is the prediction of QCD, but any
explicit evidence for a pure glueball state has never been
confirmed in the spectroscopy of isoscalar mesons.  Lattice QCD
studies \cite{Lattice} suggest the mass of lightest scalar
glueball  lies at $1.5\sim1.7\mbox{GeV}$. Among the established
resonances with the quantum numbers to be a scalar glueball, the
three mesons, $f_0(1370)$, $f_0(1500)$ and $f_0(1710)$, are the
most natural candidates. Actually, it is likely that they are the
mixtures of $\bar qq$ and glueball. Different mixing mechanisms
for these mesons were proposed in the literature
\cite{MOglue,Closeglue,LW,GGLF,CCL}. In different mixing
mechanisms, $f_0(1500)$ is described differently which will affect
the production in $B$ decays. In the future, the $B$ decay
experimental data can help us to specify the right mixing. In the
following, we assume the $f_0(1500)$ meson is dominated by the
quarkonium content, i.e. $|f_0(1500)\ra=\cos\theta |\bar ss\ra
+\sin\theta |\bar nn\ra$ and leave the contribution from glueball
content for future study.

\subsection{Decay constants and Light-Cone Distribution Amplitudes}

In two-quark picture, the decay constants for scalar meson $S$ are
defined by: \be \langle S(p)|\bar q_2\gamma_\mu
q_1|0\ra&=&f_Sp_\mu, \,\,\,\, \langle S(p)|\bar q_2
q_1|0\ra=m_S\bar {f_S}.\en Due to the charge conjugation
invariance, the neutral scalar mesons $f_0(980)$ and $f_0(1500)$
cannot be produced by the vector current. Thus $f_S=0$. Taking the
mixing into account, the above definition is changed to: \be
\langle f_0^n|\bar uu|0\ra=\frac{1}{\sqrt 2}m_{f_0}\tilde
f^n_{f_0},\,\,\,\, \langle f_0^n|\bar ss|0\ra=m_{f_0}\tilde
f^s_{f_0}. \en Using the QCD sum rules method, the decay constants
$\tilde f_{f_0}^n$ and $\tilde f_{f_0}^s$ of $f_0(980)$ have been
determined separately but with no great difference \cite{CYf0K}.
So the assumption of $\tilde f_{f_0}^n=\tilde f_{f_0}^s$  works
well. In the following, we will denote them as $\bar f_{f_0}$.

The twist-2 and twist-3 light-cone distribution amplitudes (LCDAs)
for different components of $f_0$ are defined by: \be \langle
f^{(n,s)}_0(p)|\bar q(z)_l q(0)_j|0\rangle
&=&\frac{1}{\sqrt{2N_c}}\int^1_0dxe^{ixp\cdot
z}\{p\sl\Phi^{(n,s)}_{f_0}(x)
+m_{f_0}\Phi^{(n,s)S}_{f_0}(x)\non\\
&&+m_{f_0}(n\sl_+n\sl_--1)\Phi^{(n,s)T}_{f_0}(x)\}_{jl},\label{LCDA}
\en where $n_+$ and $n_-$ are light-like vectors:
$n_+=(1,0,0_T),n_-=(0,1,0_T)$. The normalization can be related to
the decay constants: \be \int^1_0 dx\Phi^{(n,s)}_{f_0}(x)=\int^1_0
dx\Phi^{(n,s)T}_{f_0}(x)=0,\,\,\,\,\,\,\,\int^1_0
dx\Phi^{(n,s)S}_{f_0}(x)=\frac{\bar f_f}{2\sqrt{2N_c}}.\en In the
following, we assume the $SU(3)$ symmetry and use the same LCDAs
for $\bar ss$ and $\bar nn$. The twist-2 LCDA can be expanded in
the Gegenbauer polynomials: \be
\Phi_f(x,\mu)&=&\frac{1}{2\sqrt{2N_c}}\bar
f_f(\mu)6x(1-x)\sum_{m=1}^\infty B_m(\mu)C^{3/2}_m(2x-1). \en The
decay constants and the Gegenbauer moments $B_m(\mu)$  for twist-2
distribution amplitude have been studied in \cite{CYf0K,
CCYscalar} using the QCD sum rules approach:
\begin{equation}\begin{array}{lll}
 {\mbox {Scenario I:}}&\bar f_{f_0(980)}= (0.37\pm0.02)\mbox
{ GeV},&  \bar
f_{f_0(1500)}=-(0.255\pm0.03)\mbox { GeV}, \\
  &B_1(980)=-0.78\pm0.08,& B_3(980)=0.02\pm0.07, \\
&B_1(1500)=0.80\pm0.40, & B_3(1500)=-1.32\pm0.14,   \\
{\mbox {Scenario II:}} & \bar f_{f_0(1500)}=(0.49\pm0.05)\mbox { GeV},\\
&B_1(1500)=-0.48\pm0.11,& B_3(1500)=-0.37\pm0.20, \label{ff}
\end{array}\end{equation}
 where the
values for Gegenbauer moments are taken at scale $\mu=1 \mbox{
GeV}$. As for the explicit form of the Gegenbauer moments for the
twist-3 distribution amplitudes $\Phi_f^s$ and $\Phi_f^T$, they
have not been studied in the literature, so we adopt the
asymptotic form:
\be
\Phi^S_f&=& \frac{1}{2\sqrt {2N_c}}\bar
f_f,\,\,\,\,\,\,\,\Phi_f^T= \frac{1}{2\sqrt {2N_c}}\bar f_f(1-2x).
\en

In the previous PQCD study on $B\to f_0(980)K$ \cite{Chenf0K1},
the author neglected the twist-2 contribution but only used  the
twist-3 distribution amplitude $\Phi^S_f(x)$ and proposed the
following form:
\begin{eqnarray}
 \Phi^S_f(x)=\frac{\bar
f}{2\sqrt{2N_c}}\{3(1-2x)^2+\xi
(1-2x)^2[C^{3/2}_2(1-2x)-3]+1.8C^{1/2}_4(1-2x)\},
\end{eqnarray}
with $C^{1/2}_4(y)=(35y^4-30y^2+3)/8$, $C^{3/2}_2(y)=3/2(5y^2-1)$.
The decay constant $\bar f_f=0.2 \mbox { GeV} $ was used which is
close to the earlier QCD sum rules study: $\bar f_f=0.18\pm0.015
\mbox{ GeV}$ \cite{FP}. The parameter $\xi$ was chosen as:
$\xi=0.3-0.5$. While in Ref. \cite{Chenf0K2}, the twist-2
distribution amplitude was also included:
\begin{eqnarray}
\Phi_f(x)=\frac{\bar
f_f}{2\sqrt{2N_c}}G[6x(1-x)C^{3/2}_1(1-2x)],\end{eqnarray} where
$G\sim1.11$ obtained from the corresponding value in $a_0(980)$
given by \cite{DH}. But the decay constant $\bar f_f=0.2 \mbox {
GeV} $ is much smaller than the recent QCD sum rules results in
Eq.(\ref{ff}). The reason of the difference is that the scale
dependence of $\bar f_f$ and the radiative corrections to the
quark loops in the operator product expansion series is taken into
account in Ref. \cite{CYf0K,CCYscalar}. The larger decay constant
can surely enhance the branching ratio and we expect a much larger
branching ratio for the decay $B\to f_0(980)K$ also in PQCD
approach.

\section{Calculation of decays in PQCD approach and discussions} \label{results}

In the standard model, the effective weak Hamiltonian mediating
flavor-changing neutral current transitions of the type $b\to s $
has the form:
 \be
 {\cal H}_{eff}=&&{G_F\over \sqrt 2}\Big[\sum\limits_{p=u,c}V_{pb}V^*_{ps}\Big(
 C_1O_1^p+C_2O_2^p\Big)-V_{tb}V^*_{ts}\sum\limits_{i=3}^{10,7\gamma,8g}
 C_iO_i\Big],
 \en
where the explicit form of the operator $O_i$ and the
corresponding Wilson coefficient $C_i$ can be found in Ref.
\cite{Buras}. $V_{p(t)b}$, $V_{p(t)s}$ are the
Cabibbo-Kobayashi-Maskawa (CKM) matrix elements.

In the effective Hamiltonian, the degrees of freedom heavier than
$b$ quark mass $m_b$ scale is included in the Wilson coefficients
which can be calculated using the perturbation theory. Then the
left task is to calculate the operators' matrix elements between
the $B$ meson state and the final states, which suffers large
uncertainties. Nevertheless, the problem becomes tractable if
factorization becomes applicable. The PQCD approach is one of the
standard factorization approaches in hadronic B decay studies
\cite{PQCD}. In this approach, we apply the $k_T$ resummation to
kill the end-point singularities and threshold resummation to
smear the double logarithmic divergence from the weak corrections,
which results in the Sudakov form factor $S$ and the jet function
$J$ respectively. Then the decay amplitude can be factorized into
the convolution of the wave functions and the hard kernel in the
following form: \be {\cal A}=\Phi_B\otimes H \otimes J \otimes
S\otimes \Phi_{M_1}\otimes \Phi_{M_2}.\en The hard part $H$ can be
calculated perturbatively, while the LCDAs $\Phi_B$, $\Phi_{M_1}$
and $\Phi_{M_2}$, although non-perturbative in nature, are
universal for all modes. They can be determined by other well
measured decay channels to make predictions here. For example, the
corresponding light-cone distribution amplitudes of $B$ and $K$
mesons are well constrained by the $B\to K \pi$ and $B\to \pi\pi$
decays\cite{PQCD}.


\begin{table}\caption{Input parameters used in the numerical calculation}
\begin{center}
\begin{tabular}{c |cc}
\hline \hline
 Masses &$m_{f_0(980)}=0.98 \mbox{ GeV}$   &$ m_0^K=1.7 \mbox{ GeV}$, \\ \
 & $m_{f_0(1500)}=1.5
 {\mbox{ GeV}}$  & $ M_B = 5.28 \mbox{ GeV}$\\
 \hline
  Decay constants &$f_B = 0.19 \mbox{ GeV}$  & $ f_{K} = 0.16
 \mbox{ GeV}$\\
 \hline
Life Times &$\tau_{B^\pm}=1.671\times 10^{-12}\mbox{ s}$ &
$\tau_{B^0}=1.536\times 10^{-12}\mbox{ s}$\\
 \hline
$CKM$ &$V_{tb}=0.9997$ & $V_{ts}=-0.04$,\\
 &$V_{us}=0.2196$ & $V_{ub}=0.00367e^{-i60^{\circ}}$\\
\hline \hline
\end{tabular}\label{para}
\end{center}
\end{table}

The leading order Feynman diagrams are given in Fig.~\ref{ss} and
\ref{nn} for the $\bar ss$ and $\bar nn$ component of $f_0$
respectively. The decay rates of $B\to f_0K$ can be expressed as:
\begin{eqnarray}
\Gamma=\frac{G_F^2}{32\pi m_B}|A^{(-)}|^2(1-r_f^2),\end{eqnarray}
 in which $r_f=m_f/m_B$. $A$ is the decay amplitude of $\bar B^0\to f_0\bar K^0$ and $A^-$ is the decay
amplitude $B^-\to f_0 K^-$. $A^{(-)}$ can be written as
\begin{eqnarray}
A^{(-)}=A^{(-)}_{\bar
ss}\times\cos\theta+\frac{1}{\sqrt2}A^{(-)}_{\bar
nn}\times\sin\theta,\end{eqnarray}
 with \be A_{\bar ss}
&=&-V_{tb}V_{ts}^*\bigg [F_{e}^{SP}(a_6-\frac{1}{2}a_8)+{\cal
M}_{e}^{LL}(C_3+C_4-\frac{1}{2}C_9-\frac{1}{2}C_{10})+{\cal
M}_e^{LR}(C_5-\frac{1}{2}C_7)\nonumber\\
 &&\;\;\;\;+{\cal
 M}_e^{SP}(C_6-\frac{1}{2}C_8)+F_{a}^{LL}(a_4-\frac{1}{2}a_{10})
 +F_{a}^{SP}(a_6-\frac{1}{2}a_{8})\nonumber\\
 &&\;\;\;\;+{\cal M}_{a}^{LL}(C_3-\frac{1}{2}C_9)+{\cal M}_a^{LR}(C_5-\frac{1}{2}C_7)\bigg]\;,\\
 A_{\bar nn}&=&V_{ub}V_{us}^*{\cal M}_{e}^{LL}(C_2)-V_{tb}V_{ts}^*
\bigg[M_{e}^{LL}(2C_4+\frac{1}{2}C_{10})+M_{e}^{SP}(2C_6+\frac{1}{2}C_8)\nonumber\\
&&\;\;\;\;
+F_{e}^{LL\prime}(a_4-\frac{1}{2}a_{10})+F_e^{SP\prime}(a_6-\frac{1}{2}a_8)+{\cal
M}_{e}^{LL\prime}(C_3-\frac{1}{2}C_9)+{\cal
M}_e^{LR\prime}(C_5-\frac{1}{2}C_7)\nonumber\\
&&\;\;\;\;
+F_{a}^{LL\prime}(a_4-\frac{1}{2}a_{10})+F_a^{SP\prime}(a_6-\frac{1}{2}a_8)
+{\cal M}_{a}^{LL\prime}(C_3-\frac{1}{2}C_9)+{\cal
M}_a^{LR\prime}(C_5-\frac{1}{2}C_7)\bigg],\\
 A^{-}_{\bar ss} &=&
V_{ub}V_{us}^*\bigg[F_{a}^{LL}(a_1)+{\cal
M}^{LL}_{a}(C_1)\bigg]-V_{tb}V_{ts}^*\bigg
[F_{e}^{SP}(a_6-\frac{1}{2}a_8)+{\cal
M}_{e}^{LL}(C_3+C_4-\frac{1}{2}C_9-\frac{1}{2}C_{10})\nonumber\\
 &&\;\;\;\;+{\cal
M}_e^{LR}(C_5-\frac{1}{2}C_7)+{\cal
 M}_e^{SP}(C_6-\frac{1}{2}C_8)+F_{a}^{LL}(a_4+a_{10})
 +F_{a}^{SP}(a_6+a_{8})\nonumber\\
 &&\;\;\;\;+{\cal M}_{a}^{LL}(C_3+C_9)+{\cal M}_a^{LR}(C_5+C_7)\bigg]\;,\\
 A^-_{\bar
nn}&=&V_{ub}V_{us}^*\bigg[{\cal
M}_{e}^{LL}(C_2)+F_e^{LL\prime}(a_1)+{\cal M}_e^{LL\prime}(C_1)
+F_a^{LL\prime}(a_1)+{\cal M}_a^{LL\prime}(C_1)\bigg]\nonumber\\
&&\;\;-V_{tb}V_{ts}^*
\bigg[M_{e}^{LL}(2C_4+\frac{1}{2}C_{10})+M_{e}^{SP}(2C_6+\frac{1}{2}C_8)\nonumber\\
&&\;\;\;\;
+F_{e}^{LL\prime}(a_4+a_{10})+F_e^{SP\prime}(a_6+a_8)+{\cal
M}_{e}^{LL\prime}(C_3+C_9)+{\cal
M}_e^{LR\prime}(C_5+C_7)\nonumber\\
&&\;\;\;\; +F_{a}^{LL\prime}(a_4+a_{10})+F_a^{SP\prime}(a_6+a_8)
+{\cal M}_{a}^{LL\prime}(C_3+C_9)+{\cal
M}_a^{LR\prime}(C_5+C_7)\bigg],\en where the combinations of the
Wilson coefficients are defined as usual \cite{AKL}:
\begin{eqnarray}
a_1= C_2+C_1/3, & a_3= C_3+C_4/3,~a_5= C_5+C_6/3,~a_7=
C_7+C_8/3,~a_9= C_9+C_{10}/3,\\
a_2= C_1+C_2/3, & a_4= C_4+C_3/3,~a_6= C_6+C_5/3,~a_8=
C_8+C_7/3,~a_{10}= C_{10}+C_{9}/3.
\end{eqnarray}
The explicit amplitudes, for the factorizable $f_0$-emission
contribution $F_e$ (the first two diagrams in the first line of
Fig.~\ref{ss} and the first two diagrams in the third line of
Fig.~\ref{nn}) and nonfacorizable contribution ${\cal M}_e$ (the
last two diagrams in the first line of Fig.~\ref{ss} and the last
two diagrams in the third line of Fig.~\ref{nn}), the factorizable
annihilation $F_a$ (the first two diagrams in the second line of
Fig.~\ref{ss} and the first two diagrams in the second line of
Fig.~\ref{nn}) and non-factorizable annihilation contribution ${\cal
M}_a$ (the last two diagrams in the second line of Fig.~\ref{ss} and
the last two diagrams in the second line of Fig.~\ref{nn}), for the
factorizable $K$-emission contribution $F'_e$ (the first two
diagrams in the first line of Fig.~\ref{nn}) and nonfacorizable
contribution ${\cal M}_e$ (the last two diagrams in the first line
of Fig.~\ref{nn}),  are given in the appendix~\ref{formulae}.

For the numerical calculation, we list the input parameters in
Table~\ref{para}.

\subsection{Branching ratios}

At first, we give the results of the form factor $F_0^{\bar B^0\to
f_0(\bar nn)}$ at maximally recoiling:
\begin{eqnarray}
F_0^{B\to f_0(980)}&=&0.47,\\
F_0^{B\to f_0(1500)}&=&-0.39 \;\;\;{\mbox {Scenario I}}, \\
F_0^{B\to f_0(1500)}&=& 0.86\;\;\;{\mbox {Scenario II}}.
\end{eqnarray}
These form factors are large, because the decay constants of the
scalar mesons are very large. The minus sign of the $B\to
f_0(1500)$ form factor in scenario I arises from the decay
constant of $f_0(1500)$.

If $f_0(980)$ is purely composed of $\bar ss$, the branching
ratios of $B\to f_0(980)K$ are: \be {\cal B}(\bar B^0\to
f_0(980)\bar K^0)&=&(22^{+2+2+1}_{-2-2-0})\times 10^{-6},\\
{\cal B}( B^-\to f_0(980) K^-)&=&(24^{+3+3+1}_{-2-2-0})\times
10^{-6},
\end{eqnarray}
where the uncertainties are from the decay constant of $f_0(980)$,
the Gegenbauer moments $B_1$ and $B_3$. If $f_0(980)$ is purely
composed of $\bar nn$, the branching ratios for $B\to f_0(980) K$
are: \be
 {\cal B}(\bar B^0\to
f_0(980)\bar K^0)&=&(28^{+3+6+2}_{-3-5-2})\times 10^{-6},\\
{\cal B}( B^-\to f_0(980) K^-)&=&(34^{+3+5+2}_{-3-6-3})\times
10^{-6},
\end{eqnarray}
 where
the uncertainties are from the same quantities as above. Our
results are  larger than   the earlier PQCD results
\cite{Chenf0K1}, with the branching ratios for purely $\bar ss$
component
 \be {\cal
B}(\bar B^0\to
f_0(980)\bar K^0)&=&1.39\times 10^{-6},\\
{\cal B}( B^-\to f_0(980) K^-)&=&1.57\times 10^{-6},
\end{eqnarray}
 and for
purely  $\bar nn$ component, ${\cal B}(\bar B^0\to f_0(980)\bar
K^0)\simeq {\cal B}( B^-\to f_0(980) K^-) \simeq 5\times 10^{-6}$.

\begin{table}\caption{ Decay amplitudes for $\bar B^0\to f_0(980)\bar
K^0$ ($\times 10^{-2} \mbox {GeV}^3$), where  ``\cite{Chenf0K1}''
denotes the results using the LCDAs proposed in \cite{Chenf0K1},
``This work'' denotes the results using Gegenbauer moments of
twist-2 distribution amplitude from QCD sum rules and asymptotic
form of twist-3 distribution amplitudes.}
\begin{center}
\begin{tabular}{c c|c|c|c|c}
\hline \hline $\bar ss$& & $F_e$ & $M_e$ & $F_a$ & $M_a$ \\
 \hline \cite{Chenf0K1}  &   &$6.2$&$ \sim0$&$-2.6-2.1 i$&$0.04-0.06i$ \\
\hline
This work &    &11&$-4.0-5.0i$&$0.9-8.3i$&$0.31+0.49i$ \\
\hline \hline

$\bar nn$& & $M_e^T$ &  $M_e$ & $F_a$ & $M_a$  \\
 \hline \cite{Chenf0K1}&    &0&0&$-3.1-3.1i$&$-0.42+0.13i$\\
\hline
This work &    &$52+55i$&$-10 -13 i$& $1.0-9.9i$&$-0.38-0.18i$\\

\hline \hline $\bar nn$&& $F_{e}^\prime$ & $M_{e}^\prime$ &   &    \\
 \hline \cite{Chenf0K1} &  &$-7.9$&$-0.01-0.03i$&---  &---\\
\hline
This work &   &$-2.2$&$0.17+0.58i$ &--- &---\\
\hline
\end{tabular}\label{amp}
\end{center}
\end{table}

\begin{figure}[[hbtp]
\vspace{-0.2cm}
\begin{center}
\psfig{file=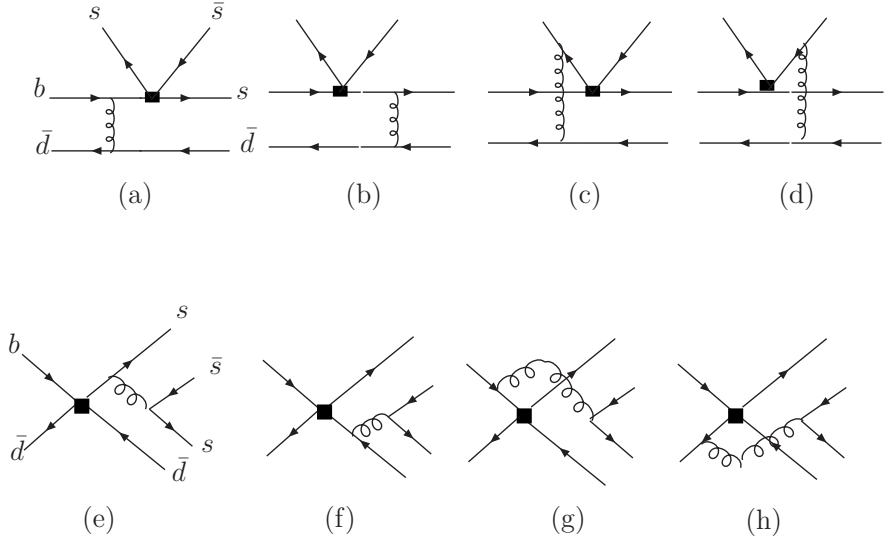,width=12.0cm,angle=0}
\end{center}
\vspace{-0.2cm} \caption{{  The leading order Feynman diagrams for
$\bar B^0\to f_0(\bar ss)\bar K^0$ decay in PQCD approach
}}\label{ss}
\end{figure}

In order to find the sources of the difference, we list the
numerical results for different topology diagrams of $\bar B^0\to
f_0(980)\bar K^0$ in Table~\ref{amp} \footnote{We reproduce the
result using the previous light-cone distribution amplitudes, with
slightly different convention:  we didn't factor  out the $-M_B^2$
for decay amplitudes; we didn't factor out the decay constant in the
factorizable amplitudes.}. In the table, $F_e$($F_a$) and $M_e$
($M_a$) denote as the $f_0$ emission (annihilation) factorizable
contributions and non-factorizable contributions from penguin
operators respectively, while $F'_{e}$ and $M'_{e}$ are the
contributions from the $K$ emission diagrams in $\bar nn$ component
of $f_0(980)$.  $M_e^T$ denotes the $f_0$ emission non-factorizable
contribution from tree operator $O_2$. From the table, we can see
that for $\bar ss$ component of $f_0(980)$ the new result is quite
larger than the previous one \cite{Chenf0K1}. The main reason is
that the new decay constant for $f_0(980)$ is twice as the previous
one. Furthermore, in Ref. \cite{Chenf0K1} the non-factorizable
contribution is very small, which is understandable from the
amplitude for this contribution: the third diagram and the fourth
diagram in the first line of Fig.~\ref{ss} cancel with each other
due to the symmetry of $x_2\to 1-x_2$ in the twist-3 distribution
amplitudes. But after including the twist-2 distribution amplitude
for $f_0(980)$, on the contrary, the two diagrams do not cancel with
each other due to the antisymmetry of the twist-2 distribution
amplitude $\Phi_f(x)$ and give a large contribution. Although the
normalization of $\Phi_f(x)$ is zero, this contribution can enhance
the total branching ratio further. In the factorizable part of
annihilation diagrams (Fig.\ref{ss}(e) and (f)), the distribution
amplitude $\Phi^S$ in \cite{Chenf0K1} can give a large contribution
both to the real part and imaginary part. But here we include the
twist-2 distribution amplitude and use the asymptotic form of
twist-3 distribution amplitudes. The real part becomes positive with
a value half of the one in \cite{Chenf0K1}, and the imaginary part
is 3 times larger. This suggests that the annihilation type
amplitude is quite sensitive to the shape of the distribution
amplitudes.

\begin{figure}[htb]
\vspace{-0.2cm}
\begin{center}
\psfig{file=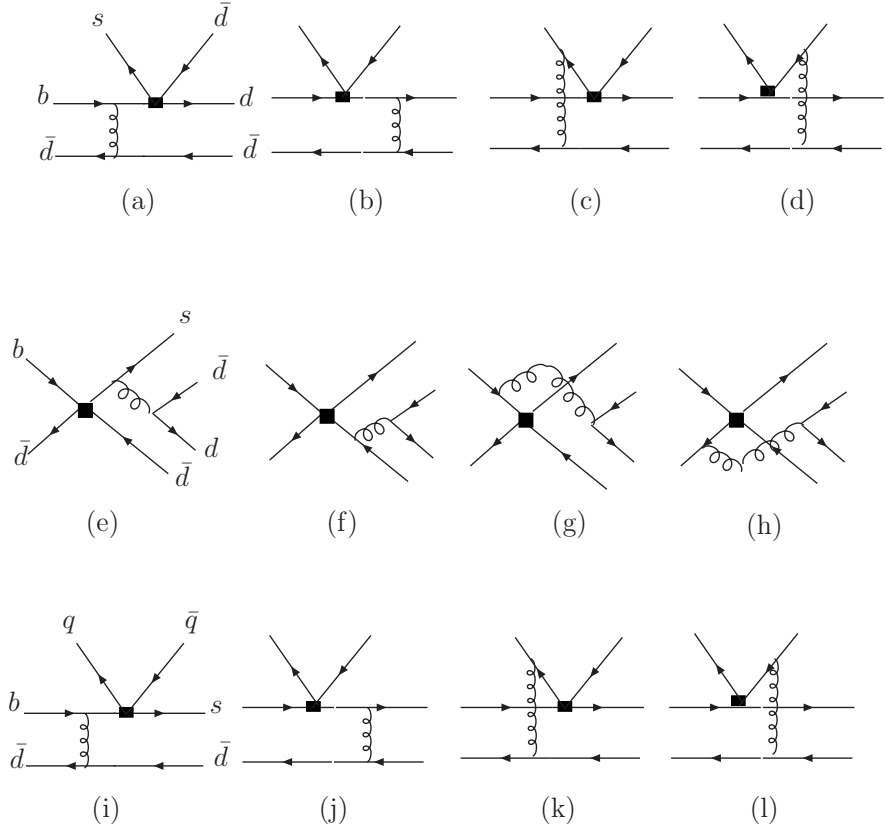,width=12.0cm,angle=0}
\end{center}
\vspace{-0.2cm} \caption{{  The leading order Feynman diagrams for
$\bar B^0\to f_0(\bar nn)\bar K^0$ decay in PQCD approach, where $q$
denotes $u$ or $d$ quark in the last row}}\label{nn}
\end{figure}

For $B\to f_0(980)(\bar nn)K$, there are two kinds of emission
diagrams: $f_0(980)$-emission (last row in Fig.\ref{nn}) and
$K$-emission (first row in Fig.\ref{nn}). In Ref. \cite{Chenf0K1},
both factorizable and non-factorizable contribution from the
$f_0(980)$-emission diagrams are zero. But after including the
twist-2 distribution amplitude $\Phi_f(x)$, the non-factorizable two
diagrams do not cancel with each other, thus can give large
contribution to branching ratio. For $K$-emission diagrams, although
we use a larger decay constant for $f_0(980)$, our result is even
smaller than the previous one. The reason is that after we include
the twist-2 distribution amplitude and use the asymptotic form for
twist-3 $\Phi_f^s$, $\Phi_f^T$, there exist cancelations between the
$(V-A)(V-A)$ operators and $(S-P)(S+P)$ operators. From the
Table~\ref{amp}, we can see that the $f_0$-emission diagrams and the
factorizable annihilation diagrams give the largest contribution,
therefore the new branching ratio is five times as the one in Ref.
\cite{Chenf0K1} for ($\bar n n$) components.

Our results are also larger than the ones   in QCD factorization
approach \cite{CCYscalar}: $ {\cal B}(B^-\to f_0K^-)\sim18\times
10^{-6}$ for $\bar ss$ component and $ {\cal B}(B^-\to
f_0K^-)\sim1\times 10^{-6}$ for $\bar nn$ component.  The main
reason is that in QCDF the $f_0$-emission nonfactorizable diagrams
are very small.

\begin{figure}[htb]
\vspace{-1.cm}
\begin{center}
\psfig{file=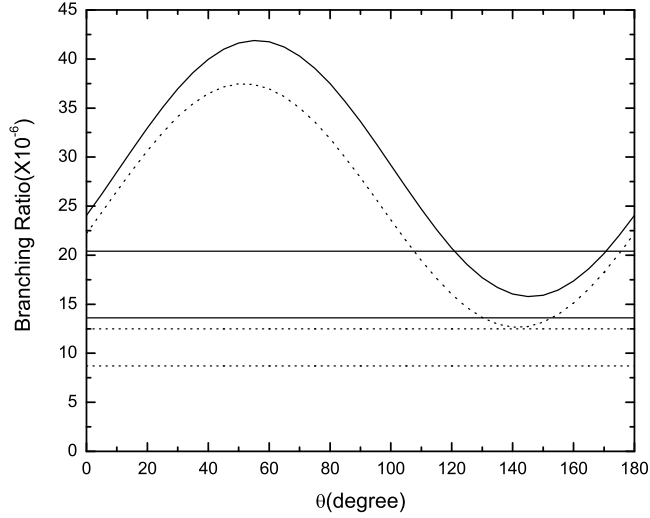,width=10.0cm,angle=0}
\end{center}
\vspace{-1.2cm} \caption{{  The dependence of the branching ratios
for $B\to f_0(980) K$ on the mixing angle $\theta$} using the inputs
derived from QCD sum rules, where the doted (solid) curve is for
$\bar B^0\to f_0\bar K^0$ ($B^-\to f_0K^-$). The horizontal band
within the doted (solid) lines shows the experimentally allowed
region of $\bar B^0\to f_0\bar K^0$ ($B^-\to f_0K^-$) within 1
$\sigma$ error.}\label{br1}
\end{figure}

In Fig.~\ref{br1}, we plot the branching ratios as functions of
the mixing angle $\theta$. Using the above mentioned range of the
mixing angle, we obtain: ${\cal B}(\bar B^0\to f_0\bar
K^0)=(32\sim36)\times 10^{-6}$, ${\cal B}( B^-\to f_0
K^-)=(35\sim39)\times 10^{-6}$ for $25^\circ<\theta<40^\circ$ and
${\cal B}(\bar B^0\to f_0\bar K^0)=(13\sim16)\times 10^{-6}$, $
{\cal B}( B^-\to f_0 K^-)=(16\sim18)\times 10^{-6} $ for
$140^\circ<\theta<165^\circ$, where only the central values of
other input parameters are used. The averaged experimental data
obtained by heavy flavor averaging group \cite{HFAG} are also
shown in Fig.~\ref{br1}: \be
{\cal B}(B^-\to f_0(980)K^-)&=&(17.1^{+3.3}_{-3.5})\times 10^{-6},\\
{\cal B}(\bar B^0\to f_0(980)\bar K^0)&=&(11.1\pm2.4)\times
10^{-6},\en where we find the PQCD approach results in the range
of $140^\circ<\theta<165^\circ$ suffice to explain the large
experimental data. Thus it does not need the existence of new
physics.

\begin{figure}[htb]
\vspace{-1.cm}
\begin{center}
\psfig{file=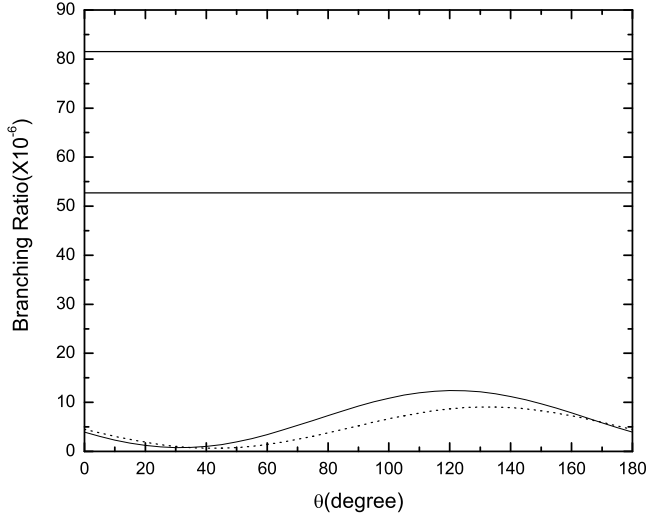,width=10.0cm,angle=0}
\end{center}
\vspace{-1.2cm} \caption{{  The dependence of the branching ratios
for $B\to f_0(1500)K$ on the mixing angle $\theta$} in scenario I,
where the doted (solid) line is for $\bar B^0\to f_0\bar K^0$
($B^-\to f_0K^-$). The horizontal band within the solid lines shows
the experimentally allowed region of $B^-\to f_0K^-$ within 1
$\sigma$ error.}\label{br2}
\end{figure}

\begin{figure}[htb]
\vspace{-1.cm}
\begin{center}
\psfig{file=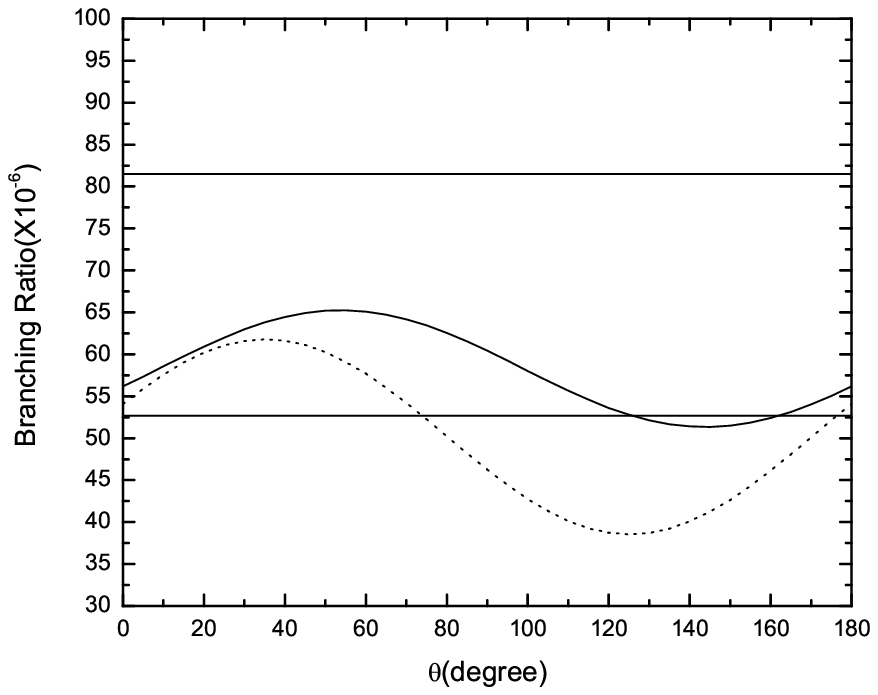,width=10.0cm,angle=0}
\end{center}
\vspace{-1.2cm} \caption{{ The dependence of the branching ratios
for $B\to f_0(1500)K$ on the mixing angle $\theta$} in scenario II,
where the doted (solid) line is for $\bar B^0\to f_0\bar K^0$
($B^-\to f_0K^-$). The horizontal band within the solid lines shows
the experimentally allowed region of $B^-\to f_0K^-$ within 1
$\sigma$ error.}\label{br3}
\end{figure}

Now we turn to $B\to f_0(1500)K$ decays. The branching ratios in
scenario I are:
 \begin{eqnarray}
  {\cal B}(\bar B^0\to
f_0(\bar ss)\bar K^0)&=&(4.5^{+1.2}_{-1.0})\times 10^{-6},\\
{\cal B}( B^-\to f_0(\bar ss)
K^-)&=&(4.0^{+1.0}_{-0.9})\times 10^{-6},\\
{\cal B}(\bar B^0\to
f_0(\bar nn)\bar K^0)&=&(5.2^{+1.3}_{-1.1})\times 10^{-6},\\
{\cal B}( B^-\to f_0(\bar nn) K^-)&=&(9.2^{+2.3}_{-2.0})\times
10^{-6},
\end{eqnarray}
 while in scenario II, the results are:
 \begin{eqnarray}
{\cal B}(\bar B^0\to
f_0(\bar ss)\bar K^0)&=&(54^{+11}_{-9})\times 10^{-6},\\
{\cal B}( B^-\to f_0(\bar ss)
K^-)&=&(56^{+12}_{-12})\times 10^{-6},\\
{\cal B}(\bar B^0\to
f_0(\bar nn)\bar K^0)&=&(46^{+10}_{-9})\times 10^{-6},\\
{\cal B}( B^-\to f_0(\bar nn) K^-)&=&(60^{+13}_{-11})\times
10^{-6},
\end{eqnarray}
where $f_0(\bar ss)$ denotes the $\bar ss$ component of
$f_0(1500)$ and similar for $f_0(\bar nn)$. The uncertainty is
coming from the decay constant of $f_0(1500)$. The decay constant
in scenario II is larger than that in scenario I, so we can get a
larger branching ratio in scenario II for both $\bar ss$ and $\bar
nn$ component, and the contributions from the two kinds of
components are very close. In scenario I, $B^-\to f_0(\bar nn)K^-$
is larger than $\bar B^0\to f_0(\bar nn)\bar K^0$, which ia from
the large tree contribution in $B^-\to f_0(\bar nn)K^-$. This also
implies that there is a large CP asymmetry in $B^\pm\to f_0(\bar
nn)K^\pm$. As discussed above, $f_0(1500)$ may be the mixture of
$\bar ss$ and $\bar nn$. So we plot the branching ratios as a
function of mixing angle in Fig.~\ref{br2} for scenario I and
Fig.~\ref{br3} for scenario II. For the comparable contribution
from  the $\bar nn$ and $\bar ss$ components, the variation range
according to the mixing angle are not very large, which can be
seen from Fig.\ref{br2} and Fig.\ref{br3}. Using the mixing
mechanism for $f_0(1500)$ in \cite{CCL}: $|f_0(1500)\ra=-0.54
|\bar nn\ra+0.84|\bar ss\ra+0.03|G\ra$ and neglecting the small
component of glueball, we get: ${\cal B}(\bar B^0\to f_0(1500)\bar
K^0)=8.7\times 10^{-6}$ and ${\cal B}( B^-\to f_0(1500)
K^-)=10\times10^{-6}$ in scenario I; ${\cal B}(\bar B^0\to
f_0(1500)\bar K^0)=42\times 10^{-6}$ and ${\cal B}( B^-\to
f_0(1500) K^-)=55\times10^{-6}$ in scenario II. With the
experimental data in \cite{BelleKpipi} and ${\cal B}(f_0(1500)\to
K^+K^-)=0.043$, we have: \be {\cal {B}}(B^-\to f_0(1500)
K^-)&=&(471.9\pm 51.3)\times 10^{-6},
 \mbox {Solution I}, \\
{\cal{B}}(B^-\to f_0(1500) K^-)&=&(67.1\pm 14.4)\times 10^{-6},
\mbox{ Solution II}, \en where we identify the resonance as
$f_0(1500)$. We can find that the second experimental solution is
more appropriate which is also consistent with \cite{MOBtoscalar}.
The experimental data of solution II for $ B^-\to f_0(1500)K^-$ is
also shown in Fig.~\ref{br2} and Fig.~\ref{br3}  if the resonance
can be viewed as $f_0(1500)$. In scenario I, our central value is
out of the experimental range with $3\sigma$, while in scenario
II, with possible mixing the experimental data can be well
explained.

In the above calculation,  the non-perturbative inputs for
$f_0(1500)$ are the decay constants and the light-cone
distribution amplitudes derived from the QCD sum rules for the
first excited scalar state or the ground scalar state with the
mass around $1.5$ GeV. Similar results can also be applied to
$f_0(1370)$ and $f_0(1710)$, if these mesons are dominated by the
quarkonium content.

\begin{figure}[htbp]
\begin{center}
\includegraphics[scale=0.65]{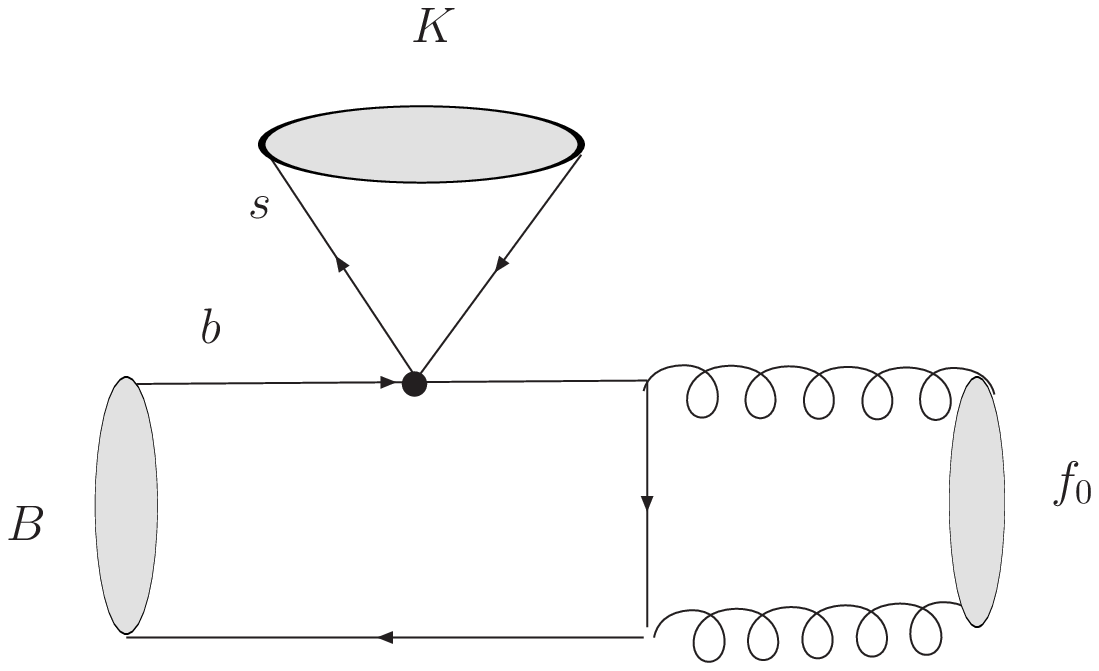}
\caption{{One of the leading order contributions to $B\to f_0
$(glueball)$ K$ in PQCD,  other diagrams obtained by attaching one
or both of the two gluons to any other quark lines are suppressed.
}}\label{glueball}
\end{center}
\end{figure}

It should be more interesting to include  the contribution from
glueball component in these decays. Thus we should consider the
typical diagram in Fig.~\ref{glueball} and other corresponding
diagrams due to the different emissions of the gluons, but the
others are suppressed as argued in \cite{QCDFsinglet}. The
complete calculation of these diagrams requires the derivation of
Sudakov form factor and jet function for a glueball state, which
is quite technical. We leave this work for future study, although
the glue component is argued to play an important role in $f_0$.

\subsection{$CP$ asymmetries}

The results of the direct $CP$ asymmetries are listed in
Table~\ref{CP}. In $\bar B^0\to f_0(\bar ss)\bar K^0$, there is no
tree contribution at the leading order, so the $CP$ asymmetry is
naturally zero. In $B^-\to f_0(\bar ss)K^-$, the tree contribution
is from the annihilation diagrams which suppressed by $1/m_B$, thus
the direct $CP$ asymmetry is small. As we have discussed, the
$f_0$-emission non-factorizable diagrams not only give large penguin
contributions but also to the tree contributions, so the $CP$
asymmetry of $B^-\to f_0(980)(\bar nn)K^-$ is large.  The direct
penguin contribution to $\bar B^0\to f_0(1500)(\bar nn)\bar K^0$ in
scenario I has some cancelation between emission diagrams and the
annihilation diagrams, thus there is a large direct $CP$ asymmetry.
But in scenario II, the annihilation penguin diagrams enhance the
emission diagrams, so the direct $CP$ asymmetry in scenario II is
rather small. In $B^-\to f_0(1500)(\bar nn)K^-$, the $CP$
asymmetries are large in both scenarios. The different sign is due
to the sign of the Gegenbauer moments.

\begin{table}
\caption{Direct $CP$ asymmetries (in units of \%):
the results in the brackets are the $CP$ asymmetries for $\bar nn$
component, the other values are for $\bar ss$
component.}\label{CP}
\begin{center}
\begin{tabular}{c|c|c|c}
   \hline \hline
   Channel & Scenario I & Scenario II &  \\
   \hline   $\bar B^0 \to f_0(980)\bar K^0$&$0(3.0)$&-\\
    $ B^- \to f_0(980)K^- $ &$-1.8(-24)$&-\\
    $\bar B^0 \to f_0(1500)\bar K^0 $&$0(24)$&$0(3.5)$\\
    $B^- \to f_0(1500)K^-$&$1.2(30)$&$-1.5(-27)$\\
   \hline\hline
\end{tabular}
   \end{center}
\end{table}

\begin{figure}[htbp]
\begin{center}
\includegraphics[scale=1.0]{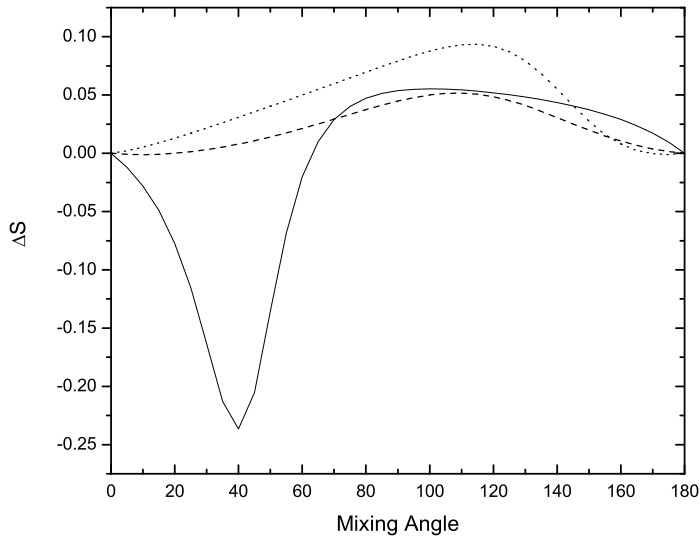}
\caption{{The $\theta$ dependence of the $\Delta S$ with the solid
line   for $B\to f_0 (1500)K_S$ in scenario I, the dashed   for
$B\to f_0(1500)K_S$ in scenario II and the doted line   for $B\to
f_0(980)K_S$. }}\label{deltas}
\end{center}
\end{figure}

Now we discuss the $CP$ violation in the neutral $B^0$ decays in
which there are both direct $CP$ asymmetry $A^{dir}_{CP}$ and
mixing-induced $CP$ asymmetry $A^{mix}_{CP}$. The time dependent
$CP$ asymmetry of $B$ decay into a $CP$ eigenstate $f$ is defined
as: \be A_{CP}(t)=A^{dir}_{CP}(B_d\to f)\cos(\Delta M
t)+A^{mix}_{CP}(B_d\to f)\sin(\Delta M t),\en with \be
A^{dir}_{CP}(B_d\to f)&=&\frac{|\lambda|^2-1}{1+|\lambda|^2},
\;\;\;
A^{mix}_{CP}(B_d\to f)=\frac{2 Im \lambda}{1+|\lambda|^2},\\
\lambda&=&\eta e^{-2i\beta}\frac{A(\bar B_d\to f)}{A(B_d\to
f),}\en where  $\eta=\pm1$ depends on the $CP$ eigenvalue of $f$,
$\Delta M$ is the mass difference of the two neutral $B$ meson
eigenstates. We can also use $C,S$ to denote the direct and
mixing-induced $CP$ asymmetry. $\beta$ is the $CKM$ angle defined
as usual \cite{PDG}. If there is no tree contribution in the
amplitude, the direct $CP$ asymmetry is zero and $\lambda$ can be
related to $e^{-2i\beta}$, so the mixing induced $CP$ asymmetry is
proportional to $\mbox{sin}(2\beta)$. For $\bar ss$ component of
$f_0$, even with the glue component, there is no tree contribution
at the leading order. This channel can be used to extract the
$CKM$ angle $\beta$. For $\bar nn$ component, there is tree
contributions. But for $B^0\to f_0(980)K^0$, the tree contribution
is small and the direct $CP$ asymmetry is only a few percent, so
this mode can serve as a possible place to extract $\beta$ even
when taking the mixing of $f_0(980)$ into account. In scenario II,
$B\to f_0(1500)K$ is similar. In scenario I, the tree contribution
to $B\to f_0(1500)(\bar nn)K$ is relatively large and the direct
$CP$ asymmetry is roughly $24\%$. Using $\mbox{sin}(2\beta)=0.687$
\cite{PDG}, the mixing-induced $CP$ asymmetries of $B\to f_0(\bar
nn)K$ are:
 \begin{eqnarray}
 A_{CP}^{mix}(\bar B^0\to f_0(980)K_{S})&=&-0.608,\\
  A_{CP}^{mix}(\bar B^0\to f_0(1500)K_{S})&=&-0.633, \;\;\;\;\mbox{Scenario I},\\
  A_{CP}^{mix}(\bar B^0\to f_0(1500)K_{S})&=&-0.642,\;\;\;\; \mbox{Scenario
  II}.
 \end{eqnarray}
They are not far away from $-{\mbox{sin}}(2\beta)=-0.687$. On the
experimental side, the parameter $\Delta S$ is often used:
\begin{eqnarray}
\Delta S=A_{CP}^{mix}+\mbox{sin} (2\beta).
\end{eqnarray}
 As we
have discussed, the direct $CP$ asymmetry in  $B^0\to f_0(\bar
ss)K_S$ decay vanishes thus the parameter $\Delta S$  is zero.  For
$\bar nn$ component, the extra tree contribution makes $\Delta S$
deviate from 0. If new physics can induce the $b\to s$ transitions
and has a different phase, it would also give a non-zero value to
this parameter. Precise measurement of this parameter and the
theoretical calculation in the standard model can also help us to
probe new physics. Our results for these three channels in standard
model  are:
\begin{eqnarray}
\Delta S(\bar B^0\to f_0(980)K_{S})&=&0.079,\\
 \Delta S(\bar B^0\to f_0(1500)K_{S})&=&0.054, \;\;\;\;\mbox{Scenario I},\\
 \Delta S(\bar B^0\to f_0(1500)K_{S})&=&0.045,\;\;\;\; \mbox{Scenario
 II}.
\end{eqnarray}
The dependence of $\Delta S$ on the mixing angle $\theta$ of $f_0$
is also plotted in Fig.~\ref{deltas}. From this figure, we can see
that for $B^0\to f_0(980)K_S$  and $B^0\to f_0(1500)K_S$ decay in
scenario II, there is not much deviation from $\Delta S=0$. But
there are large deviations from $\Delta S=0$ for the interference of
$\bar ss$ and $\bar nn$ component of $f_0(1500)$ in scenario I
around $\theta=40^\circ$.

\subsection{Theoretical Uncertainties}

In our calculation, one of the uncertainties is from the scalar
meson decay constants. These uncertainties can give sizable
effects on the branching ratio, but the $CP$ asymmetries are less
sensitive to these parameters. There are other uncertainty sources
for both branching ratios and $CP$ asymmetries:

\begin{itemize}
\item  The twist-3 distribution amplitudes of the scalar mesons
are taken as the asymptotic form for lack of better results from
non-perturbative methods, this may give large uncertainties. These
distribution amplitudes needs to be studied in future work.

\item The Gegenbauer moments $B_1$ and $B_3$ for twist-2 LCDAs of
$f_0(980)$ and $f_0(1500)$ have sizable uncertainties. For
example, the large uncertainty of $B_1$ in scenario I and $B_3$ in
scenario II may change the results sizably, these parameters
should be constrained in future.

\item The uncertainties, from the light pseudoscalar meson and B
meson wave functions, $\Lambda_{QCD}$ and other renormalization
group parameters, $O(1/M_B)$ corrections and the sub-leading
component of $B$ meson distribution amplitude, have been
systemically studied extensively in \cite{kurimoto}: the uncertainty
from the factorization scale choice is within $10\%$; the results
vary by $10-30\%$ by changing the parameter in the wave functions.

\item The sub-leading order contributions in PQCD approach have
also been neglected in the calculation, which were calculated in
Refs. \cite{subleadingPQCD} for $B\to \pi\pi,\pi K$, etc. These
corrections can change the penguin dominated processes, for
example, the quark loops and magnetic-penguin correction decrease
the branching ratio of $B\to \pi K$ by about $20\%$. We expect the
similar size of uncertainty in $B\to f_0K$ decays, since they are
also dominated by the penguin operators.

\item Besides, the decay amplitude suffer other power corrections
which are non-perturbative in nature: the long distance
re-scattering effect. This effect could be phenomenologically
included in the final-state interactions \cite{FSI}. But we need
more data to determine whether it is important in $B\to SP$
decays.
\end{itemize}


\section{Conclusion}\label{summary}

In this work, we re-analyze the exclusive decays $B\to f_0(980) K$
in perturbative QCD approach by identifying $f_0(980)$ as the
composition of $\bar ss$ and $\bar nn=(\bar uu+\bar dd)/\sqrt2$.
The $B\to f_0(1500)K$ is also analyzed in PQCD approach. Our main
results are as follows:
\begin{itemize}

\item Using the decay constants and light-cone distribution
amplitude derived from QCD sum rules, we find that the PQCD
results can also explain the large experimental data which agrees
with results from QCDF.

\item{The non-factorizable $f_0$-emission type diagrams can give
large contributions, although the normalization of the twist-2
distribution amplitude for $f_0$ is zero.}

\item The $B\to f_0(1500)K$ decay is studied under the assumption
of quarkonium dominance in two scenarios. The $\bar nn=(\bar
uu+\bar dd)/\sqrt2$ and $\bar ss$ can give similar contributions.
In scenario II, the branching ratios are large which can
accommodate with the second solution of the experimental data, but
in scenario I, the predicted branching ratio is smaller than the
experimental data.

\item The  calculation of $B\to f_0(1500)K$ decays can also be
applied to $f_0(1370)$ and $f_0(1710)$, if these mesons are
dominated by the quarkonium content.

\item The mixing-induced $CP$ asymmetries are not far away from
${\mbox{sin}}(2\beta)=0.687$ for $B^0\to f_0(980)K_{S,L}$ and
$B^0\to f_0(1500)K_{S,L}$. Thus  these channels can provide
possible places to extract the $CKM$ angle $\beta$.

\end{itemize}

\section*{Acknowledgment}
This work was supported by the National Science Foundation of
China. We would like to thank C.H. Chen, H.Y. Cheng, C.K. Chua, D.
Dujmic, Z.T. Wei and Q. Zhao for helpful discussions and comments.
W.W. would like to thank Y.M. Wang and F.Q. Wu for debugging the
program.

\begin{appendix}
\section{factorization formulae}\label{formulae}

In this appendix, we give the factorization formulae involved in
the decay amplitudes. In the formulae, we choose the momentum
fraction at the anti-quark, thus we should use $\Phi_f(1-x)$,
$\Phi_f^S(1-x)=\Phi^S_f(x)$ and $\Phi_f^T(1-x)=-\Phi_f^T(x)$ for
$f_0$. But for simplicity, we will use $\Phi_f(x)$ to denote
$\Phi_f(1-x)$ in the formulae. It is similar for the pseudoscalar
meson $K$.

In  each Feynman diagram, there may be three different kinds of
operators: $(V-A)(V-A)$, $(V-A)(V+A)$, $(S-P)(S+P)$ (arising from
the Fierz transformation of $(V-A)(V+A)$ operators). We will use the
indices $LL$, $LR$ and $SP$ to characterize the different kinds of
the operators.

If $f_0$ is emitted, the factorization formulae for the emission
type diagrams are:
\begin{eqnarray}
 F^{SP}_{e}(a_i) &=& -16 \pi C_F m_B^4 r_{f}
{\overline{f_{f}}}\int_0^1 dx_1 dx_3 \int_0^{\infty} b_1db_1\,
b_3db_3\, \Phi_B(x_1,b_1)
\nonumber \\
& &\times \bigg\{ \left[ \Phi_{K}^A(x_3)+r_{K} x_3 \left(
\Phi_{K}^P(x_3) - \Phi_{K}^T(x_3) \right) +2 r_{K} \Phi_{K}^P(x_2)
 \right]\nonumber\\
&&\;\;\;\;  \times E_{ei}(t) h_{e}(x_1,x_3,b_1,b_3) + 2 r_{K}
 \Phi_K^P(x_3)
E_{ei}(t') h_{e}(x_3,x_1,b_3,b_1) \bigg\}\;,
\end{eqnarray}
 for the factorizable diagrams, i.e. Fig.~\ref{ss}(a),(b)
and Fig.~\ref{nn} (i),(j), and
\begin{eqnarray}
 {\cal M}^{LL}_{e}(a_i) &=& -32 \pi C_Fm_B^4
/\sqrt{2N_C}\int_0^1 dx_1dx_2dx_3 \int_0^{\infty} b_1 db_1\, b_2
db_2\,\Phi_B(x_1,b_1) \Phi_{f}(x_2)
\non \\
&&\bigg\{\left[(x_2-1)\Phi_{K}(x_3)+r_{K}x_3(\Phi_{K}^P(x_3)-\Phi_{K}^T(x_3))\right]
E'_{ei}(t) h_n(x_1,1-x_2,x_3,b_1,b_2)
\nonumber \\
&& +\left[(x_2+x_3)\Phi_{K}(x_3) -r_{K} x_3 \left(
\Phi_{K}^P(x_3)+\Phi_{K}^T(x_3)\right)
       \right]
E'_{ei}(t') h_n(x_1,x_2,x_3,b_1,b_2) \bigg\}\;,
\end{eqnarray}
\begin{eqnarray}
{\cal M}^{LR}_{e}(a_i) &=& 32 \pi C_Fm_B^4r_{f}
/\sqrt{2N_C}\int_0^1 dx_1dx_2dx_3 \int_0^{\infty} b_1 db_1\, b_2
db_2\,\Phi_B(x_1,b_1)
\nonumber \\
& &\times \bigg\{ E'_{ei}(t)
h_n(x_1,1-x_2,x_3,b_1,b_2)\times\left[ (x_2-1)\Phi_{K}^A(x_3)
\left( \Phi_{f}^S(x_2) + \Phi_{f}^T(x_2) \right)
\right.\nonumber \\
& &\;\;\;\;\;\;+r_{K}(x_2-1)\left( \Phi_{K}^P(x_3)-\Phi_{K}^T(x_3)
\right) \left( \Phi_{f}^S(x_2)+ \Phi_{f}^T(x_2)
\right)\nonumber\\
& &\;\;\;\;\;\;\left. - r_{K} x_3 \left(
\Phi_{K}^P(x_3)+\Phi_{K}^T(x_3) \right) \left( \Phi_{f}^S(x_2)-
\Phi_{f}^T(x_2) \right)\right]
\nonumber \\
& & \;\;\;\;+  E'_{ei}(t') h_n(x_1,x_2,x_3,b_1,b_2)\times \left[
x_2\Phi_{K}^A(x_3) \left( \Phi_{f}^S(x_2) - \Phi_{f}^T(x_2)
\right) \right.\nonumber \\
&&\;\;\;\;\; +r_{K}x_2\left( \Phi_{K}^P(x_3)-\Phi_{K}^T(x_3)
\right) \left( \Phi_{f}^S(x_2)- \Phi_{f}^T(x_2)
\right)\nonumber\\
& &\;\;\;\;\;\;\left. + r_{K} x_3 \left(
\Phi_{K}^P(x_3)+\Phi_{K}^T(x_3) \right) \left( \Phi_{f}^S(x_2)+
\Phi_{f}^T(x_2) \right)\right]\bigg\}\;,
\end{eqnarray}
\begin{eqnarray}
{\cal M}^{SP}_{e}(a_i) &=& -32 \pi C_Fm_B^4 /\sqrt{2N_C}\int_0^1
dx_1dx_2dx_3 \int_0^{\infty} b_1 db_1\, b_2
db_2\,\Phi_B(x_1,b_1)\Phi_f(x_2)\bigg\{E'_{ei}(t)
\nonumber \\
& &\times \left[ (x_2-x_3-1)\Phi_{K}^A(x_3) +r_{K}x_3\left(
\Phi_{K}^P(x_3)+\Phi_{K}^T(x_3) \right)\right]
h_n(x_1,1-x_2,x_3,b_1,b_2)
\nonumber \\
& & \;\;\;+ \left[x_2\Phi_{K}^A(x_3)-r_K
x_3\left(\Phi_K^P(x_3)-\Phi_K^T(x_3)\right)\right] E'_{ei}(t')
h_n(x_1,x_2,x_3,b_1,b_2)\bigg\}\;,
\end{eqnarray}
for the nonfactorizable diagrams, i.e. Fig.~\ref{ss} (c),(d) and
Fig.~\ref{nn} (k),(l). While if $K$ is emitted, the formulae are:
\begin{eqnarray}
F^{LL\prime}_{e}(a_i) &=& 8 \pi C_F m_B^4 f_{K}\int_0^1 dx_1 dx_2
\int_0^{\infty} b_1db_1\, b_2db_2\, \Phi_B(x_1,b_1)
\nonumber \\
& & \times \bigg\{ \left[ (1+x_2)\Phi_{f}(x_2)-r_{f}(1-2x_2)
\left( \Phi_{f}^S(x_2)+\Phi_{f}^T(x_2) \right) \right] E_{ei}(t)
h_{e}(x_1,x_2,b_1,b_2)
\nonumber\\
& &\;\;\;\;\;\; -2r_{f} \Phi_{f}^S({x_2})E_{ei}(t')
h_{e}(x_2,x_1,b_2,b_1) \bigg\} \;,\end{eqnarray}
\begin{eqnarray}
 F^{SP\prime}_{e}(a)&=&16 \pi C_F
m_B^4 f_{K}r_K\int_0^1 dx_1 dx_2 \int_0^{\infty} b_1db_1\,
b_2db_2\, \Phi_B(x_1,b_1)
\nonumber \\
& &\times \bigg\{ -\left[ \Phi_{f}(x_2)+r_{f} \left(
x_2\Phi_{f}^T(x_2)-(x_2+2)\Phi_{f}^S(x_2) \right) \right]
E_{ei}(t) h_{e}(x_1,x_2,b_1,b_2)
\nonumber\\
& &\;\;\;\;\;\; +2r_{f} \Phi_{f}^S({x_2})E_{ei}(t')
h_{e}(x_2,x_1,b_2,b_1) \bigg\} \;,\end{eqnarray}
\begin{eqnarray}
{\cal M}^{LL\prime}_{e}(a_i) &=& -32 \pi C_Fm_B^4
/\sqrt{2N_C}\int_0^1 dx_1dx_2dx_3 \int_0^{\infty} b_1 db_1\, b_2
db_2\,\Phi_B(x_1,b_1) \Phi_{k}^A(x_3)
\nonumber \\
&&\times
 \bigg\{\left[(x_3-1)\Phi_{f}(x_2)-r_{f}x_2(\Phi_{f}^S(x_2)-\Phi_{f}^T(x_2))\right]
E'_{ei}(t) h_n(x_1,1-x_3,x_2,b_1,b_3)
\nonumber \\
&&+\left[ (x_2+x_3)\Phi_{f}(x_2) +r_{f} x_2
\left(\Phi_{f}^S(x_2)+\Phi_{f}^T(x_2)\right)
       \right]
E'_{ei}(t') h_n(x_1,x_3,x_2,b_1,b_3) \bigg\}\;,
\end{eqnarray}
\begin{eqnarray}
{\cal M}^{LR\prime}_{e}(a_i) &=& 32 \pi C_Fm_B^4
/\sqrt{2N_C}r_K\int_0^1 dx_1dx_2dx_3 \int_0^{\infty} b_1 db_1\,
b_3 db_3\,\Phi_B(x_1,b_1)
\nonumber \\
&&\times
\bigg\{E'_{ei}(t) h_n(x_1,1-x_3,x_2,b_1,b_3)\times
\left[(x_3-1)\Phi_f(x_2)(\Phi^P_K(x_3)+\Phi^T_K(x_3))\right.\nonumber\\
&&\;\;\;+r_f\Phi_{f}^T(x_2)
\left((x_2+x_3-1)\Phi^P_K(x_3)+(-x_2+x_3-1)\Phi_K^T(x_3)\right)
\nonumber\\
 && \left.\;\;\;+r_{f}\Phi_{f}^S(x_2)\left((x_2-x_3+1)\Phi_K^P(x_3)-(x_2+x_3-1)\Phi_K^T(x_3)
 \right)\right]
\nonumber\\
&&-\left[x_3\Phi_{f}(x_2)\left(\Phi^T_K(x_3)-\Phi_K^P(x_3)\right)+r_fx_3
\left(\Phi_f^S(x_2)-\Phi_f^T(x_2)\right)
\left(\Phi_K^P(x_3)-\Phi_K^T(x_3)\right)\right.\nonumber\\
&& \left.\;+r_fx_2
\left(\Phi_f^S(x_2)+\Phi_f^T(x_2)\right)\left(\Phi_K^P(x_3)+\Phi_K^T(x_3)\right)\right]
E'_{ei}(t') h_n(x_1,x_3,x_2,b_1,b_3) \bigg\}\;.
\end{eqnarray}

In the annihilation diagrams, if $f_0$ is the upper meson (in the
heavy $b$ quark side), the factorization formulae are:
\begin{eqnarray}
F^{LL}_{a}(a_i) &=& 8 \pi C_F m_B^4f_B \int_0^1 dx_2 dx_3
\int_0^{\infty} b_2db_2\, b_3db_3\
\nonumber \\
&& \times \bigg\{[(x_3-1)\Phi_{K}^A(x_3)\Phi_{f}(x_2) - 2r_{K}
r_{f}(x_3-2) \Phi_{K}^P(x_3)\Phi_{f}^S(x_2) +2r_{K} r_{f}x_3
\Phi_{K}^T(x_3)\Phi_{f}^S(x_2) ]
\nonumber \\
&&\;\;\;\times E_{ai}(t) h_{a}(x_2,1-x_3, b_2, b_3)
\nonumber\\
&&
 \;\;\;+ [x_2\Phi_{K}^A(x_3)\Phi_{f}(x_2)-2r_{K}
r_{f}\Phi_{K}^P(x_3)((x_2+1)\Phi_{f}^S(x_2)+(x_2-1)\Phi_{f}^T)]\nonumber\\
&& \;\;\;\;\times E_{ai}(t') h_{a}(1-x_3,x_2, b_3, b_2)\bigg\},
\en
\be \nonumber F^{SP}_{a}(a_i) &=& -16 \pi C_F m_B^4f_B \int_0^1 dx_2
dx_3 \int_0^{\infty} b_2db_2\, b_3db_3\,\bigg\{E_{ai}(t)
h_{a}(x_2,1-x_3, b_2, b_3)\times
\\ \nonumber
 &&
\left[r_{K}(x_3-1)\Phi_{f}(x_2)\left(\Phi_{K}^P(x_3)+\Phi_{K}^T(x_3)\right)
+2r_{f}\Phi_{K}(x_3)\Phi_{f}^S(x_2)\right]-h_{a}(1-x_3,x_2, b_3,
b_2)
\\ &&\times
\left[2r_{K}\Phi_{K}^P(x_3) \Phi_{f}(x_2)+r_{f}x_2\Phi^A_{K}(x_3)
\left(\Phi_{f}^T(x_2)-\Phi_{f}^S(x_2)\right)\right]E_{ai}(t')
\bigg\}\;, \en

\be {\cal M}^{LL}_{a}(a_i) &=& 32\pi C_Fm_B^4 /\sqrt{2N_C}\int_0^1
dx_1dx_2dx_3 \int_0^{\infty} b_1 db_1\, b_2 db_2\,\Phi_B(x_1,b_1)
\nonumber \\
& &\times \bigg\{ E'_{ai}(t) h_{na}(x_1,x_2,x_3,b_1,b_2)\left[
-x_2\Phi_{K}^A(x_3) \Phi_{f}(x_2)
\right.\nonumber \\
& &+r_{K}r_{f} \Phi_{f}^T(x_2) \left(
(x_2+x_3-1)\Phi_{K}^P(x_3)+(-x_2+x_3+1) \Phi_{K}^T(x_3)
\right)\nonumber\\
& &\;\;\;\;\;\;\left. + r_{K}r_{f} \Phi_{f}^S(x_2) \left(
(x_2-x_3+3)\Phi_{K}^P(x_3)-(x_2+x_3-1) \Phi_{K}^T(x_3)
\right)\nonumber\right]
\nonumber \\
& & \;\;\;\;-  E'_{ai}(t')
h'_{na}(x_1,x_2,x_3,b_1,b_2)\times\left[ (x_3-1)\Phi_{K}^A(x_3)
\Phi_{f}(x_2)
\right.\nonumber \\
& &\;\;\;\;\;\;+r_{K}r_{f} \Phi_{K}^P(x_3) \left(
(x_2-x_3+1)\Phi_{f}^S(x_2)-(x_2+x_3-1) \Phi_{f}^T(x_2)
\right)\nonumber\\
& &\;\;\;\;\;\;\left. + r_{K}r_{f} \Phi_{K}^T(x_3) \left(
(x_2+x_3-1)\Phi_{f}^S(x_2)-(1+x_2-x_3) \Phi_{f}^T(x_2)
\right)\right] \bigg\}\;, \en

\begin{eqnarray}
 {\cal M}^{LR}_{a}(a_i) &=& 32 \pi C_Fm_B^4 /\sqrt{2N_C}\int_0^1
dx_1dx_2dx_3 \int_0^{\infty} b_1 db_1\, b_2
db_2\,\Phi_B(x_1,b_1)\times
\nonumber \\
&& \bigg\{ \left[r_{K} (1+x_3)\Phi_{f}(x_2)
(\Phi_{K}^T(x_3)-\Phi_{K}^P(x_3))
+r_{f}(x_2-2)\Phi_{K}(x_3)(\Phi_{f}^S(x_2)+\Phi_{f}^T(x_2))\right]\nonumber \\
&&\;\;\;\;\times E'_{ai}(t) h_{na}(x_1,x_2,x_3,b_1,b_2)
\nonumber \\
&& \;\;\;- \left[r_{K} (x_3-1)\Phi_{f}(x_2)
(\Phi_{K}^T(x_3)-\Phi_{k}^P(x_3))+r_{f}x_2\Phi_{K}(x_3)(\Phi_{f}^S(x_2)+\Phi_{f}^T(x_2))
\right]\nonumber \\
&&\;\;\;\;\times  E'_{ai}(t') h'_{na}(x_1,x_2,x_3,b_1,b_2)
\bigg\}\;.
\end{eqnarray}
If $f_0$ is the lower one, 
\begin{eqnarray}
F^{LL\prime}_{a}(a_i) &=& 8 \pi C_F m_B^4f_B \int_0^1 dx_2 dx_3
\int_0^{\infty} b_2db_2\, b_3db_3
\bigg\{\left[(x_2-1)\Phi_{K}^A(x_3)\Phi_{f}(x_2)\right.
\nonumber \\
&& \left. + 2r_{K} r_{f}(x_2-2) \Phi_{K}^P(x_3)\Phi_{f}^S(x_2)
-2r_{K} r_{f}x_2 \Phi_{K}^P(x_3)\Phi_{f}^T(x_2) \right]
\nonumber \\
&&\;\;\;\;\times E_{ai}(t) h_{a}(x_3,1-x_2, b_3, b_2)
+E_{ai}(t') h_{a}(1-x_2,x_3, b_2, b_3)\times\nonumber\\
&&
  \left[x_3\Phi_{K}^A(x_3)\Phi_{f}(x_2)+2r_{K}
r_{f}\Phi_{f}^S(x_2)\left((x_3+1)\Phi_{K}^P(x_3)+(x_3-1)\Phi_{K}^T(x_3)\right)\right]
  \bigg\},
\end{eqnarray}
\begin{eqnarray}
F^{SP\prime}_{a}(a_i) &=& 16 \pi C_F m_B^4f_B \int_0^1 dx_2 dx_3
\int_0^{\infty} b_2db_2\, b_3db_3\, \bigg\{ E_{ai}(t)
h_{a}(x_3,1-x_2,
b_2, b_3)\nonumber\\
&&
\times\left[r_{f}(x_2-1)\Phi_{K}^A(x_3)\left(\Phi_{f}^S(x_2)+\Phi_{f}^T(x_2)\right)
-2r_{K}\Phi_{K}^P(x_3)\Phi_{f}(x_2)\right]\nonumber\\
 &&  - \left[2r_{f}\Phi^A_{K}(x_3)
\Phi_{f}^S(x_2)+r_{K}x_3\Phi_{f}(x_2)
\left(\Phi_{K}^P(x_3)-\Phi_{K}^T(x_3)\right)\right]\nonumber\\
 &&\times E_{ai}(t') h_{a}(1-x_2,x_3,
b_2, b_3)\bigg\},
\end{eqnarray}
\begin{eqnarray}
 {\cal M}^{LL\prime}_{a}(a_i) &=& -32\pi C_F m_B^4 /\sqrt{2N_C}\int_0^1
dx_1dx_2dx_3 \int_0^{\infty} b_1 db_1\, b_3 db_3\,\Phi_B(x_1,b_1)
\nonumber \\
& &\times \bigg\{ E'_{ai}(t) h_{na}(x_1,x_3,x_2,b_1,b_3)\left[
x_3\Phi_{K}^A(x_3) \Phi_{f}(x_2)
\right.\nonumber \\
& &+r_{K}r_{f} \Phi_{f}^T(x_2) \left(
(x_2-x_3+1)\Phi_{K}^T(x_3)-(x_2+x_3-1) \Phi_{K}^P(x_3)
\right)\nonumber\\
& & \left. + r_{K}r_{f} \Phi_{f}^S(x_2) \left(
(-x_2+x_3+3)\Phi_{K}^P(x_3)+(x_2+x_3-1) \Phi_{K}^T(x_3) \right)\
\right]
\nonumber \\
& &  +  E'_{ai}(t') h'_{na}(x_1,x_3,x_2,b_1,b_3)\left[
(x_2-1)\Phi_{K}^A(x_3) \Phi_{f}(x_2)
\right.\nonumber \\
& & +r_{K}r_{f} \Phi_{f}^T(x_2) \left(
(-x_2+x_3+1)\Phi_{K}^T(x_3)-(x_2+x_3-1) \Phi_{K}^P(x_3)
\right)\nonumber\\
& & \left. + r_{K}r_{f} \Phi_{f}^S(x_2) \left(
(x_2-x_3-1)\Phi_{K}^P(x_3)+(x_2+x_3-1) \Phi_{K}^T(x_3) \right)
\right] \bigg\}\;,
\end{eqnarray}
\begin{eqnarray}
{\cal M}^{LR\prime}_{a}(a_i) &=& 32 \pi C_Fm_B^4
/\sqrt{2N_C}\int_0^1 dx_1dx_2dx_3 \int_0^{\infty} b_1 db_1\, b_3
db_3\,\Phi_B(x_1,b_1)\bigg\{E'_{ai}(t)
\nonumber \\
&&\times  \left[r_{f} (x_2+1)\Phi_{K}^A(x_3)
(\Phi_{f}^S(x_2)-\Phi_{f}^T(x_2))
+r_{K}(x_3-2)\Phi_{f}(x_2)(\Phi_{K}^P(x_3)+\Phi_{K}^T(x_3))\right]\nonumber \\
&& \;\;\;\;\times  h_{na}(x_1,x_3,x_2,b_1,b_3)-E'_{ai}(t')
h'_{na}(x_1,x_3,x_2,b_1,b_3)\times
\nonumber \\
&&  \left[r_{f} (x_2-1)\Phi_{K}^A(x_3)
(\Phi_{f}^S(x_3)-\Phi_{f}^T(x_3))+r_{K}x_3\Phi_{f}(x_2)(\Phi_{K}^P(x_3)+\Phi_{K}^T(x_3))
\right]  \bigg\}\;.
\end{eqnarray}
In the above formulae, $r_f=m_f/m_B$ and $r_K=m_K^0/m_B$, $m_0^K$
is the chiral scale parameter. The function $E$ are defined as:
\begin{eqnarray}
E_{ei}(t)&=&\alpha_s(t)\,a_i(t)\exp[-S_B(t)-S_{3}(t)],\\
E'_{ei}(t)&=&\alpha_s(t)\,a_i(t)\exp[-S_B(t)-S_{2}(t)-S_{3}(t)]|_{b_3=b_1},\\
E_{ai}(t)&=&\alpha_s(t)\,a_i(t)\exp[-S_2(t)-S_{3}(t)],\\
E'_{ai}(t)&=&\alpha_s(t)\,a_i(t)\exp[-S_B(t)-S_{2}(t)-S_{3}(t)]|_{b_3=b_2},
\end{eqnarray}
where $\alpha_s$ is the strong coupling constant and $a_i$ is the
corresponding Wilson coefficient, $S$ is the Sudakov form factor.
In our numerical analysis, we use the one-loop expression for the
strong coupling constant; we use  $c=0.3$ for the parameter in the
jet function. The explicit form of $h$ and $S$ have been given in
\cite{PQCD}.
 \end{appendix}


\end{document}